\documentclass[twocolumn,iop,apjl]{emulateapj}
\usepackage[utf8]{inputenc}
\usepackage{xcolor}
\usepackage{pdfcolmk}
\usepackage{units}
\usepackage{amsbsy}
\usepackage{amssymb}
\usepackage{graphicx}
\usepackage{esint}
\PassOptionsToPackage{normalem}{ulem}
\usepackage{ulem}

\makeatletter

\providecolor{lyxadded}{rgb}{0,0,1}
\providecolor{lyxdeleted}{rgb}{1,0,0}

\journalinfo{ \textit{submitted}} 
\slugcomment{submitted}

\usepackage{bm}

\@ifundefined{textcolor}{}{%
 \definecolor{BLACK}{gray}{0}
 \definecolor{WHITE}{gray}{1}
 \definecolor{RED}{rgb}{1,0,0}
 \definecolor{GREEN}{rgb}{0,1,0}
 \definecolor{BLUE}{rgb}{0,0,1}
 \definecolor{CYAN}{cmyk}{1,0,0,0}
 \definecolor{MAGENTA}{cmyk}{0,1,0,0}
 \definecolor{YELLOW}{cmyk}{0,0,1,0}
}

\usepackage{amsfonts}\usepackage{bm}\usepackage{color}

\def \beq {\begin{equation}}
\def \eeq {\end{equation}}
\def \bea {\begin{eqnarray}}
\def \eea {\end{eqnarray}}
\def \bfig {\begin{figure}}
\def \efig {\end{figure}}

\newcommand{\commento}[1]{{{{ #1 }}}}

\makeatother

\begin{document}

\title{Relaxation Processes in Solar Wind Turbulence}

\author{S. Servidio$^{1},$ C. Gurgiolo$^{2}$, V. Carbone$^{1}$, and M.
L. Goldstein$^{3}$}

\affil{$^{1}$Dipartimento di Fisica, Università della Calabria, I-87030
Rende (CS), Italy; sergio.servidio@fis.unical.it\\
 $^{2}$Bitterroot Basic Research, Hamilton, Montana, USA\\
 $^{3}$Heliospheric Physics Laboratory, Code 672, NASA Goddard Space
Flight Center, Greenbelt, MD,}

\begin{abstract}
Based on global conservation principles, magnetohydrodynamic (MHD)
relaxation theory predicts the existence of several equilibria, such
as the Taylor state or global dynamic alignment. These states are
generally viewed as very long-time and large-scale equilibria, which
emerge only after the termination of the turbulent cascade. As suggested
by hydrodynamics and by recent MHD numerical simulations, relaxation
processes can occur during the turbulent cascade that will manifest themselves
as local patches of equilibrium-like configurations. Using multi-spacecraft
analysis techniques in conjunction with Cluster data, we compute the
current density and flow vorticity and for the first time demonstrate
that these localized relaxation events are observed in the solar wind. Such 
events have important consequences for the statistics of
plasma turbulence.
\end{abstract}

\section{Introduction}

Intermittent turbulence and long-time relaxation processes represent
two central features of magnetohydrodynamics (MHD). Turbulence, ubiquitous
in fluids and in plasmas, cannot be fully described by simple models.
In particular, a random-phase modeling of turbulence cannot capture 
the bursty and intermittent nature of the field gradients. Observations suggest that turbulent 
plasmas are characterized by high kurtosis of field fluctuations,
multifractal behavior of the high-order structure functions, and other manifestations of 
intermittency that coexist with the cascade process \citep{Sorris-Valvo99,Burlaga01,Bruno13}.
\commento{
   Kurtosis and other higher-order moments were studied also in the 
   context of compressible MHD with application to molecular clouds 
   and to the diffuse interstellar medium \citep{BurkhartEA09,BurkhartEA10}.
}
On a parallel path, the MHD theory of relaxation processes has been
very successful in describing commonly observed features such as Taylor
states (minimum energy states conserving magnetic helicity), selective
decay, global dynamic alignment, and helical 
dynamo \citep{TaylorJB74,Montgomery78,Matthaeus82,Stribling91,Carbone92,MGoldstein95,Mininni02,Mininni05}.
Relaxation is generally viewed as a slow consequence of multiple global
ideal conservation principles leading as a final state to large-scale
equilibria. 
\commento{
   Very little has been said about a possible link between
   intermittent turbulence, relaxation processes, and other critical features 
   of MHD, such as the spectral anisotropy commonly observed in the presence of 
   a mean magnetic field \citep{Shebalin83,GoldreichSridhar95,Cho00}.
}
The idea that local equilibrium patches are embedded in turbulence forms the
basis of the present work.

It has been observed that in Navier-Stokes (NS) turbulence relaxation 
emerges quickly and locally, with important consequences for turbulence 
statistics \citep{Pelz85,Kraichnan88,Kerr87}.  
In hydrodynamic experiments, in fact, it is commonly observed that the cascade 
produces states in which the  velocity (\textbf{v}) and 
vorticity ($\mathrm{\mathbf{\boldsymbol{\omega}}}=\boldsymbol{\nabla}\times\mathbf{v}$) fields are strongly 
aligned \citep{Tsinober92}. The latter is due to 
the conservation of the global kinetic helicity 
$H_{v}=\langle\mathbf{v}\cdot\mathbf{\boldsymbol{\omega}}\rangle$, where
$\langle \dots \rangle$ denotes spatial averages. 
This alignment effect causes a suppression 
of the nonlinear interactions to levels much lower than in the case of Gaussian 
field modeling, suggesting that these local relaxation structures may be a crucial 
ingredient of the cascade \citep{MoffattHK84}.

Here we establish multiple links between relaxation,
turbulence and intermittency in an astrophysical plasma.  
An observational precursor of the present work can be found in \citet{Osman11}, 
where, 
\commento{using datasets from} 
the Advanced Composition Explorer spacecraft, 
it was shown that the solar wind velocity tends to align locally
with the magnetic field.
In a recent laboratory plasma experiment \citep{Gray13},
it has also been observed that fluctuations in a plasma embedded in
a magnetized cylinder exhibit a tendency to generate local force-free
states. Using Cluster data and inspired by recent numerical experiments
\citep{Matthaeus08,Servidio08} we explore the possibility that several
local and simultaneous relaxation processes occur in the turbulent
solar wind. 
These consist of magnetic fields parallel to the current density (force-free), aligned velocity
and magnetic fluctuations (Alfv\'enic), and correlated current and vorticity fields.

\section{Relaxation Processes in MHD}
MHD is a leading model in the study of plasma turbulence at low frequencies
and at spatial scales larger than the Larmor radius. The (incompressible)
MHD equations are written as
\begin{eqnarray}
\frac{\partial\mathrm{\mathrm{\mathbf{v}}}}{\partial t} & = & \mathbf{v}\times\boldsymbol{\omega}+\mathbf{j}\times\mathbf{b}-\boldsymbol{\nabla}P^{*}+\nu\boldsymbol{\nabla}^{2}\mathrm{\mathbf{v}},\nonumber \\
\frac{\partial\mathrm{\mathbf{b}}}{\partial t} & = & \boldsymbol{\nabla}\times(\mathbf{v}\times\mathbf{b})+\eta\boldsymbol{\nabla}^{2}\mathbf{b},\label{eq:momm0}
\end{eqnarray}
where \textbf{b} and $\mathrm{\mathbf{j}=\boldsymbol{\nabla}\times\mathbf{b}}$
are the magnetic and the current density field respectively,
$\boldsymbol{\nabla}\cdot\mathbf{b}=0$, $\nu$ and $\mu$ are viscous coefficients, 
and the pressure $P^{*}=p+(\mathrm v^{2}/2)$ satisfies the  
incompressibility condition, namely  $\boldsymbol{\nabla}\cdot\mathbf{v}=0$.
(We confine our attention here to incompressible turbulence both to
simplify the discussion and because turbulence in the solar wind is
in a nearly incompressible state, see, e.g., \citealp{MarschTu90}.)

MHD relaxation involves both \textbf{v} and \textbf{b} fields and
the family of equilibria known as Beltrami fields that can be obtained
from variational principles \citep{Montgomery78}. 
In particular, minimum energy states for MHD are obtained from the 
variational problem \citep{Stribling91}
\begin{equation}
\delta\int[(|\mathbf{v}|^{2}+|\mathbf{b}|^{2})-\alpha\mathbf{v}\cdot\mathbf{b}-
\gamma\mathbf{a}\cdot\mathbf{b}]d^{3}x=0\,\mathrm{,}
\label{eq:intvar}
\end{equation}
where $\alpha$ and $\gamma$ are constants, and \textbf{a} is the potential vector such that 
$\mathbf{b}=\boldsymbol{\nabla}\times\mathbf{a}$.
Holding constant cross helicity $H_{c}=\langle\mathbf{v}\cdot\mathbf{b}\rangle$ and magnetic 
helicity $H_{m}=\langle\mathbf{a}\cdot\mathbf{b}\rangle$, Equation (\ref{eq:intvar}) 
minimizes the energy $E=\langle|{\bf v}|^{2}+|{\bf b}|^{2}\rangle$. 
The solutions of Equation (\ref{eq:intvar}) can be summarized as
\begin{equation}
\mathbf{v}=c_{1}\mathbf{b}=c_{2}\mathbf{j}=c_{3}\boldsymbol{\omega},
\label{eq:bel3}
\end{equation}
where $c_{j}$ are combinations of $\alpha$ and $\gamma$. Note that the above 
Beltrami solutions include the Taylor force-free state, 
$\mathbf{j}\sim\mathbf{b}$ \citep{TaylorJB74}, and 
the  Alfvén solutions, with $\mathbf{v}\sim\mathbf{b}$.  
Global Alfv\'enic states are sometimes observed in the solar 
wind \citep{Belcher71,Dobrowolny80,Osman11}.
Steady, driven MHD also shows
alignment at small scales \citep{Mason06,Matthaeus08}. It is crucial
to note that solutions given by Equation (\ref{eq:bel3}) cancel the
nonlinear terms in Equations (\ref{eq:momm0}), and are therefore
equilibria.

\commento{
Decaying simulations of MHD turbulence show that, after a long time, an
asymptotic equilibrium is reached, characterized by long wavelength
states that can be force-free and/or
Alfv\'enic \citep{Montgomery78,Stribling91,Mininni05}.
}
Generally, global processes of relaxation require many characteristic 
times, after which energy is dissipated at small scales. 
Recently, \citet{Servidio08} have shown through MHD simulations that
alignment processes can also appear locally and very quickly, during the 
cascade process. 
In particular, coherent structures, characterized by the 
force-free and the Alfv\'enic states, spontaneously emerge in turbulence. 
\commento{
   This may be due to the fact that the growing time of these inertial range 
   patterns are comparable to the nonlinear times, which are 
   much smaller than the global relaxation times.
   Although these phenomena have not been numerically investigated 
   in a driven stationary case, there is supporting evidence that 
   these patches, similar to the situation in NS turbulence, cause a 
   suppression of the strength of the nonlinear interactions.
}

In analogy with previous theoretical studies \citep{Kraichnan88,Servidio08},
we investigate the relaxation principle by computing the Probability
Distribution Functions (PDFs) of the cosine-angle
\begin{equation}
\cos\theta=\frac{\mathbf{f}\cdot\mathbf{g}}{|\mathbf{f}||\mathbf{g}|},\label{eq:dot1}
\end{equation}
where $\{{\bf f}, {\bf g}\}$ represents one of \{\textbf{v}, \textbf{b}\},
\{\textbf{v}, $\boldsymbol{\omega}$\} \{\textbf{j}, \textbf{b}\} and
\{\textbf{j}, $\boldsymbol{\omega}$\}. Note that the exact Beltrami
correlations would be manifested as peaked distributions at $\cos\theta=\pm1$.
The presence of a finite cross and/or magnetic helicity will produce
a skewness in the distribution in the cosines. On the other hand,
Gaussian uncorrelated variables produce a flat distribution of $\mathrm{PDF(\cos\theta)=0.5}$.
\commento{
   The latter value is simply related to the bounded nature of the cosine-angle, 
   $-1\leq\cos\theta\leq 1$, and to the definition of PDF, 
   $\int_{-1}^{1} $PDF$(\cos\theta) d\cos\theta=1$ \citep{Matthaeus08,Servidio08}.
}
The appearance of Beltrami flows is implicitly associated with patterns where 
nonlinearity is suppressed, and where the energy cascade is 
therefore inhibited \citep{Kerr87}. Alignment is strong in these regions due to a local 
conservation of magnetic and cross-helicity. 
Note also that the presence of a very intense guide field, $\mathcal{\mathrm{B}}_{0}$,
may break the conservation of the magnetic helicity \citep{Shebalin06,Stribling91,StriblingGhosh94},
although in the case of a weak guide magnetic field 
\commento{(with respect to the level}
of fluctuations) the magnetic helicity may be considered
weakly conserved \citep{Servidio05}.

\section{Solar Wind Analyses}
To investigate MHD turbulent relaxation summarized by Equations (\ref{eq:bel3})
and (\ref{eq:dot1}) both the fields and their respective curls are
needed. Direct observations of vorticity and current density require
volumetric measurements that involve simultaneous
measurements at a minimum of four non-coplanar positions \citep{Dunlop02}.
This was not possible until the launch of Cluster. With Cluster data, measurements
of 3D properties and symmetries of the magnetic fluctuations have
been possible using data from the Flux Gate Magnetometer (FGM)
and Spatio-Temporal Analysis of Field Fluctuations (STAFF) experiments
\citep[see, e.g., ][]{Narita06,Sahraoui10}. The thermal electron
experiment on Cluster [the Plasma Electron and Current Experiment
(PEACE)] has been used to measure electron vorticity by using computed
spatial derivatives of the electron moments \citep{Gurgiolo10,Gurgiolo11}. Note that at low frequencies
(i.e., in the MHD limit) the latter represent a good measure of the
proton vorticity as well ${\boldsymbol\omega_{e}}\sim\boldsymbol\omega_{p}\equiv\boldsymbol\omega$.
This correspondence has been confirmed from previous analysis that compared
electron moments from PEACE with ion moments from the Cluster Ion Spectrometry experiment
(CIS) \citep{Gurgiolo10}. The current density has been computed using
a similar multi-spacecraft technique.  
To summarize, the present work uses data from several Cluster experiments: 
electron data come from PEACE, while data from FGM, Electric Field and Waves (EFW), 
CIS and WHISPER are used to describe local conditions, to verify the electron data, 
and to support the conclusions. 
The data cadence has been synchronized to be 4s and both field and velocity measurements
have been interpolated to the geometric center of the tetrahedron.

Data have been carefully screened to meet multiple criteria. We have
checked that the measurements are not in regions magnetically connected
to the bow shock by making sure that there is no evidence
of “return” electrons \citep[i.e., electrons that have been reflected off or
leaked through the bow shock, see, e.g.,][]{Wu84,Gosling89,Larson96}. Within
the analyzed intervals the Cluster spacecraft are in a good tetrahedral
configuration. In addition, errors such as timing and position uncertainties,
moment computation errors, inter-spacecraft calibration, and errors
in the linear approximation (used for computing the derivatives) have
been carefully checked and handled. A detailed discussion of many
of the above error sources can be found in \citet{Chanteur98,Vogt98},
and how they were approached in the context of the present analysis
is discussed in \citet{Gurgiolo10}. The error estimated on the computations
of plasma moments are of order of 6\%, while for spatial derivatives
the error is of order 15\%.

\begin{figure}
\includegraphics[scale=.45]{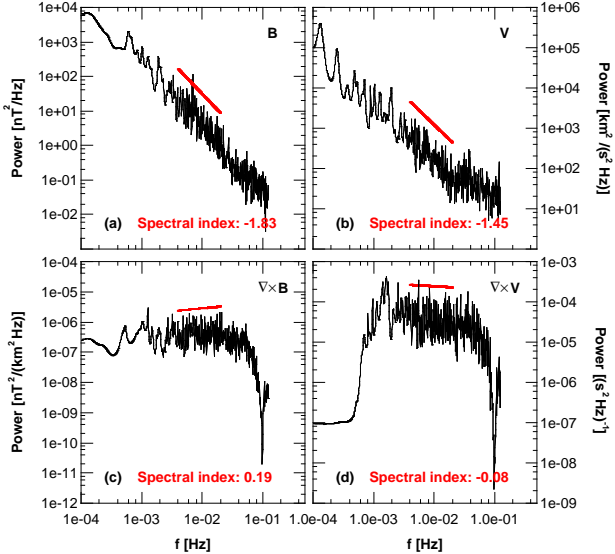}
\caption{Power spectra of the magnetic field (a), velocity (b), current density
(c) and vorticity (d), for the 2003 time-period. The current density has been obtained with
a low pass filter (\emph{f}$\mathrm{<}$0.05Hz), while the vorticity
with a bandpass filter (0.0005$\mathrm{<}$\emph{f}$\mathrm{<}$0.05Hz).
\commento{
The straight (red) lines are least square fits (which have
been offset), and the slopes are given at the bottom of the plots.
}
\label{fig:PwrSpec}}
\end{figure}

We report on two time periods, 
\commento{6 February 2003 16h10 to 18h30 UT and 6 June 2006 04h30 to 05h50 UT.} 
These time intervals meet all of the requirements listed above. Figure 1 shows the power spectra
of the magnetic field (a) and the velocity (b) for the
the 2003 time period. The higher frequencies
in the plots extend to 0.125Hz (the Nyquist frequency). 
The frequencies  correspond to timescales \commento{on the order of} the correlation time, which 
at this radial distance is approximately 1 hour. 
The magnetic field spectrum is slightly steeper than the velocity 
spectrum \citep[cf.,][]{Bruno13,Podesta07}.

Prior to computing the cosine angles in Equation (\ref{eq:dot1}) it
is necessary to precondition the data. The very low frequency components
need to be filtered out of the mean solar wind speed before the computation
of the vorticity. The presence of these low frequencies may swamp
the velocity fluctuations, especially in the radial components. This
is done by applying a high pass filter to the velocity data (a filter
frequency 0.0005 Hz) to remove the mean flow. Also, the presence of
high frequency noise in both $\mathrm{\mathbf{\nabla}\times\mathbf{v}}$
and $\mathrm{\mathrm{\mathbf{\nabla}\times\mathbf{b}}}$ may affect
gradients and act to suppress the existence of alignment effects.
The latter problem can be easily rectified by using a low pass filter
to remove frequencies above 0.04Hz. Figure 1 also shows the spectra
of the current density (c) and the vorticity (d). Note that the power
spectrum of the vortical fields has a flatter slope 
\commento{than do the}
original fields, consistent with the small scale
nature of the gradients. 
The velocity and magnetic spectra are generally consistent with the Kolmogorov expectation for
isotropic fluid turbulence, showing a wide inertial range where the
power P($k$)$\sim k^{-\nicefrac{5}{3}}$. Here $k$ represents
the wavenumber obtained using the Taylor frozen-in-flow hypothesis.
From simple dimensional arguments, since the current density Fourier
component $\mathrm{j_k\sim {\emph k}b_{k}}$, and analogously the vorticity
$\mathrm{\omega_{k}\sim {\emph k}v_{k}}$, the power spectra of the vortical
flows will scale as $k^{2}\mathrm{P({\emph k})}$. 
\commento{
   In the inertial range, therefore, these gradient fields are consistent 
   with an expectation of $\sim k^{-\nicefrac{5}{3}+2}=k^{\nicefrac{1}{3}}$.
}

\textcolor{black}{}
\begin{figure}
\textcolor{black}{\includegraphics[scale=0.45]{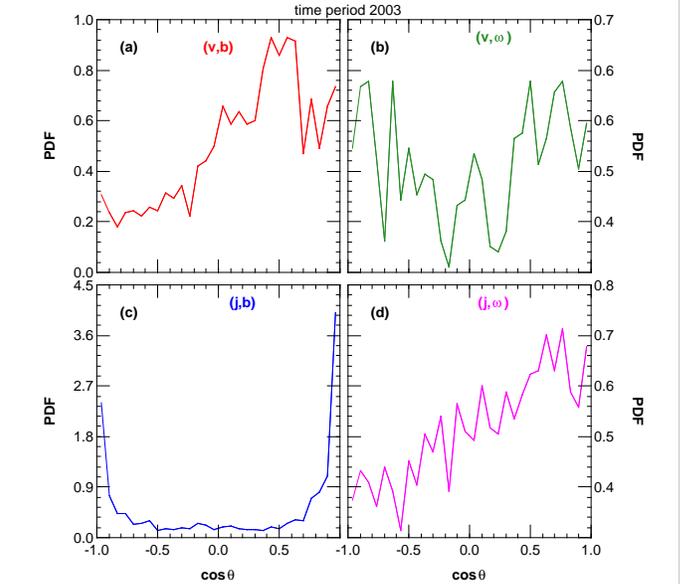}}
\textcolor{black}{\caption{\textcolor{black}{Probability Distribution Functions (PDFs) of the
cosine angle defined in }Equation\textcolor{black}{{} (\ref{eq:dot1})
for the following alignments: (a) $\{\bf{v,b}\}$, (b) $\{\bf{v,\boldsymbol\omega}\}$,
(c) $\{\bf{j,b}\}$, (d) \{{\bf j}, $\boldsymbol{\omega}$\}. 
The analysis has been performed for the 2003 dataset.\label{fig:2003cosang}}}
}
\end{figure}

To explore the possible presence of alignment patches in the turbulence,
we computed the PDFs of the cosine angles defined by Equation (\ref{eq:dot1}),
namely $\mathrm{cos\theta_{vb}}$, $\mathrm{cos\theta_{jb}}$, $\mathrm{cos\theta_{v\omega}}$,
and $\mathrm{cos\theta_{j\omega}}$, for both solar wind periods.
\commento{
   Note that the filtering procedure described before has been applied to 
   all the fields prior to the computation of each cosine-angle \citep{Kerr87}.
}
As seen in Figure 2(a), the velocity and magnetic field in
the 2003 time period exhibit a tendency to develop a highly aligned
state, with a skewed probability towards $\mathrm{cos\theta_{vb}}$
= +1. The skewness is due to the presence of a finite amount of cross
helicity in the solar wind, while the more pronounced shape is related
to the patchiness of the data. Fluctuations of the magnetic field
and velocity alignment tend to have a pronounced probability at high correlations
even if they have a lower global correlation \citep{Matthaeus08,Osman11}.
These aligned states are not present in the correlation of the velocity
with the vorticity, namely $\cos\theta_{v\omega}$. As seen in Figure
2(b), the distribution is very noisy and flat, with PDF($\mathrm{cos\theta_{v\omega}}$)
$\mathrm{\sim}$ 0.5, typical of random variables. The latter is due
to the non-conservation of kinetic helicity in MHD turbulence, which
is the main alignment in hydrodynamics \citep{Kraichnan88}.

\textcolor{black}{}
\begin{figure}
\textcolor{black}{\includegraphics[scale=0.45]{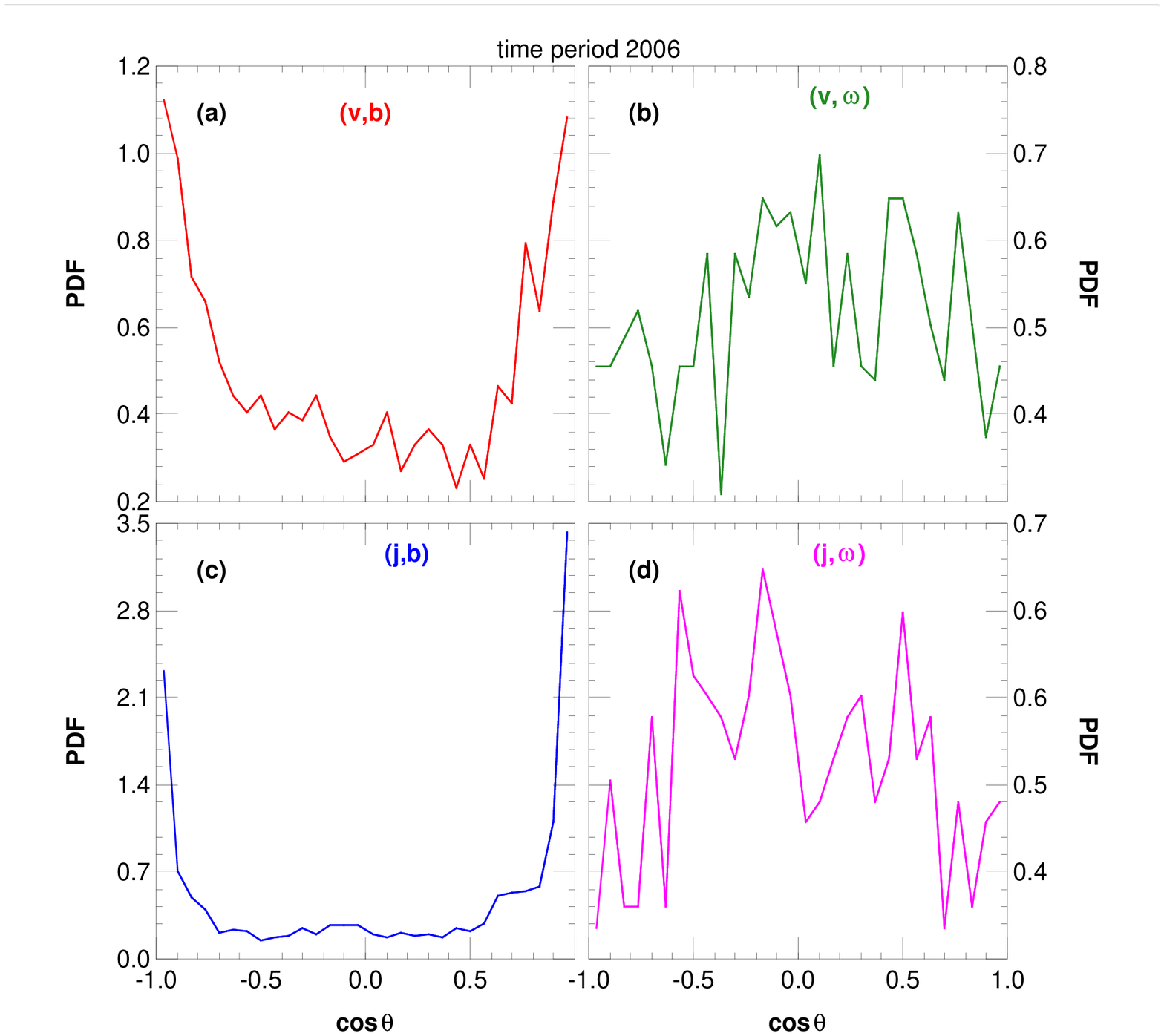}}
\textcolor{black}{\caption{\textcolor{black}{PDFs of the cosine angles for the 2006 
time interval.\label{fig:2006cosang}}}
}
\end{figure}

The strongly peaked $\mathrm{cos\theta_{jb}}$ distribution at $\mathrm{\pm1}$
in Figure \ref{fig:2003cosang}(c) indicates the presence of local
Taylor equilibria in the data that further confirms the prevalence
of local relaxation processes. Note that most of the cosine angles
are very weakly populated, and that this alignment, as in simulations,
is the most common. As can be deduced from Equation (\ref{eq:bel3}),
these Taylor states include the so-called local anisotropy \citep{Cho00}.
\commento{
   It is well known in MHD theory that the presence of a 
   mean field produces anisotropy, introducing a preferred direction 
   along the guide field \citep{Shebalin83,GoldreichSridhar95,ChoEA02,ChoLazarian03}.
}
In the latter case, the current density
is parallel to the mean field. This phenomenon, even if it can be
questioned in terms of the ergodic theorem \citep{Matthaeus12a},
can be viewed as a subclass of the Taylor states that can be 
manifest locally in solar wind turbulence.

Finally, in Figure 2(d), we plot the alignment between the current
density and the vorticity. Viewed as a consequence of the 
Alfv\'enic relaxed states this suggests that current sheets and vortex filaments
are correlated, which appears as a correlation between magnetic and
velocity field intermittency. Obviously, this state is strongly related
to the vb-correlation in Figure \ref{fig:2003cosang}(a) except that
it is much more sensitive to small scale features.

In Figure \ref{fig:2006cosang}, the same analysis has been performed
for the 2006 sample. The typical double peaked shape of the alignment
is now observed for both $\mathrm{cos\theta_{vb}}$ and $\mathrm{cos\theta_{jb}}$
as predicted in \citet{Servidio08}. 
The high probability of occurrence of Beltrami fields ($\mathrm{\mathrm{cos\theta}=\pm1}$) 
is associated  with local equilibria that emerge from turbulence via rapid 
relaxation processes  in which $\mathrm{H_{m}}$ and $\mathrm{H_{c}}$ are locally conserved. 
In contrast with hydrodynamics, in the MHD variational problem the kinetic 
helicity $\mathrm{H_{v}}$ is not conserved and therefore the 
\{\textbf{\textcolor{black}{v}}, \textbf{\textcolor{black}{$\boldsymbol{\omega}$}}\}
alignment is lost in favor of the \{\textbf{\textcolor{black}{v}}, \textbf{\textcolor{black}{b}}\}
and the \{\textbf{\textcolor{black}{j}}, \textbf{\textcolor{black}{b}}\} correlations.
\commento{
   The main difference between the results in Figure \ref{fig:2003cosang}
   and Figure \ref{fig:2006cosang} is in the distribution 
   of $\mathrm{cos\theta_{vb}}$. As we will discuss below, with respect to 2006, 
   the 2003 stream has a more pronounced and definite global cross-helicity value.
   Note also that the $\mathrm{cos\theta_{j\omega}}$ correlation is less visible
   in the 2006 time period, possibly due to both noise level and to a weaker 
   statistical convergence.
}


\begin{figure}
\begin{centering}
\includegraphics[scale=0.2]{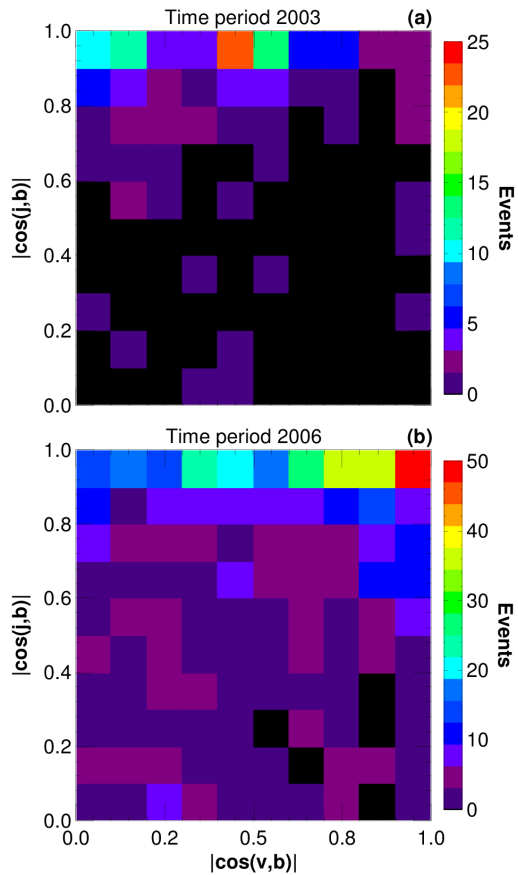}
\caption{
\commento{Joint distribution of events of $\mathrm{|cos\theta_{jb}|}$ vs.~$\mathrm{|cos\theta_{vb}|}$, 
for the 2003 (a) and 2006 (b) time periods. As it can be seen, Alfv\'enic and Taylor alignment
are correlated, revealing that Alfvénic patches are likely to be also
force-free.}
}
\label{fig:ScatterVB}
\end{centering}
\end{figure}

\commento{
   To investigate whether the local Alfv\'enic states are correlated with
   the local Taylor equilibria, we produced a two-dimensional distribution of events
   in the plane given by $\mathrm{|cos\theta_{jb}|}$ and $\mathrm{|cos\theta_{vb}|}$,
   as shown in Figure \ref{fig:ScatterVB}, for both time periods. As it can be 
   observed, especially in the 2006 case where the global cross helicity is lower, 
   the primary \textbf{v}-\textbf{b} alignment correlates with the force-free states
   showing that equilibria in turbulence follow the complex relations
   summarized in Equation (\ref{eq:bel3}).
} This effect is related to
the global relaxation processes where the final equilibrium solutions
lie on an ellipse attractor \citep{Ting86,Stribling91} as given by
the following equation 
\begin{equation}
(1-2|\sigma_{m}|)^{2}+(2\sigma_{c})^{2}=1\label{eq:ellip}
\end{equation}
where $\sigma_{c}=H_{c}/E$ and $\sigma_{m}=H_{m}/E$. 
\commento{
   These normalized helicities indicate qualitatively the skewness of the 
   cosine angles distributions. For example, for the 2003 time period, 
   we obtained $\sigma_c\sim0.7$, while, for Figure \ref{fig:2006cosang}, 
   $\sigma_c\sim-0.4$. On the contrary, we found negligible magnetic 
   helicity, consistent with expectations of inertial range 
   fluctuations \citep{Matthaeus82}.
}

\section{Conclusions and Discussion}

Using Cluster data to compute current density and flow vorticity we
have shown that local alignments of several types are observed in
the turbulent solar wind. Based on conservation of global energy,
cross helicity and magnetic helicity, MHD relaxation theory predicts
the existence of several equilibrium states, such as the Taylor magnetic
equilibrium and the global dynamic alignment. Previous theoretical
work suggested that these states, which are generally viewed as very
long-time and large-scale equilibria, appear only after the termination
of the turbulent cascade. Here we have found that these patchy relaxation
processes can coexist with the turbulent and intermittent cascade.
Even in cases where the global correlation is null, 
$H_c\simeq H_m\simeq 0$, the cosine-angles 
distribution may become concentrated near $\pm 1$. 
The presence of these equilibrium-like patterns requires that, statistically, the 
distributions of the fields, together with their gradients, become non-Gaussian. These 
results suggest that relaxation processes induce a suppression of nonlinearity in the 
solar wind. This ``cellularization'' of turbulence is not consistent with a superposition 
of random fields, 
\commento{
   and therefore involves phenomena such as intermittency and other 
   non-Gaussian features, which necessarily involves high-order correlations 
   that can be captured via multifractal analysis \citep{SheLeveque94,KowalEA07,Bruno13}.
}

It is important to note, however, that some of these effects may be
limited in the presence of very strong background magnetic fields
since the conservation of $H_m$ in
this case is broken. However in the majority of the solar wind where
the magnetic field is not that strong, the quasi-conservation of $H_m$
may allow these force-free states as observed in Figures 2(c) and 3(c). 
\commento{
   Similar features, moreover, can manifest themselves in compressible turbulence. 
   In certain regimes, indeed, the interstellar medium can be supersonic, and 
   analogous phenomena can be investigated in the context of compressible relaxation 
   theory \citep{Ghosh90}.
}

\commento{
   Longer streams of solar wind data may help to better quantify the degree of 
   local alignment, and to perform more direct 
   comparison with simulations \citep{Matthaeus08}.
}
On the basis of the present results, however, nothing can be
said about how (and where) these processes developed. They may in
principle emerge during the solar wind expansion, or at the early
stage of coronal dynamics and then slowly evolve as the the wind flows
outward.
\commento{
   A deeper investigation of the role of the mean 
   magnetic field and compressibility, as well as direct comparison with 
   simulations, will be presented in future works.
}

This research was partially supported by the “Turboplasmas” project
(Marie Curie FP7 PIRSES-2010- 269297), POR Calabria FSE 2007/2013,
and the Cluster project at the Goddard Space Flight Center.

\end{document}